\documentclass[conference,10pt,letterpaper]{IEEEtran}

\usepackage[T1]{fontenc}
\usepackage{cite}
\usepackage{amsmath,amssymb,amsfonts}
\usepackage{algorithmic}
\usepackage{graphicx}
\usepackage{textcomp}
\usepackage{xcolor}
\usepackage[hyphens]{url}

\newcommand{\sysname}{ZipFusion}

\title{ZipFusion: Fusing Lossless KV Decompression with Attention for Efficient LLM Decoding}

\author{
\IEEEauthorblockN{Yuetao Chen}
\IEEEauthorblockA{The Chinese University of Hong Kong}
\and
\IEEEauthorblockN{Peng Wang}
\IEEEauthorblockA{Unaffiliated}
\and
\IEEEauthorblockN{Hong Xu}
\IEEEauthorblockA{The Chinese University of Hong Kong}
}

\begin{document}

\maketitle

\begin{abstract}
The key-value (KV) cache places growing demands on GPU memory capacity
and bandwidth as large language models serve longer contexts and more
concurrent requests. Its footprint limits concurrency, while repeated
KV reads during attention can make decoding memory-bandwidth bound. Prior lossless
KV cache compression methods reduce transfers between devices but
restore the uncompressed KV cache in GPU memory before attention. They therefore
leave these repeated KV reads unchanged and require additional work and
memory for decompression. We present \textsc{\sysname{}}, which enables
attention to consume the losslessly compressed KV cache directly.
\sysname{} co-designs the compression format and attention kernel to
reduce the bytes read from GPU memory during attention. The kernel
reconstructs the original values directly in registers using simple
bit operations, avoiding a separate decompression pass and easing the
memory-bandwidth bound.
\sysname{} also reassigns memory freed by compression across
layers to store additional KV cache. Its memory manager
reserves space for ongoing generation and makes the remaining
capacity available to new requests. We implement
\sysname{} on the top of FlashInfer and integrate
it with SGLang.
On an NVIDIA H100, \sysname{} achieves about $\boldsymbol{1.25\times}$ speedup over
FlashInfer in decode attention. With 8K context, it improves model decode throughput by up to
\textbf{26.3\%} over uncompressed SGLang under the same GPU memory budget.
At each method's largest supported batch, it also delivers up to
$\boldsymbol{2.76\times}$ the throughput of the best-performing lossless
KV cache compression baseline.
\end{abstract}

\section{Introduction}
\label{sec:intro}

Serving large language models (LLMs) with long contexts requires substantial
GPU memory capacity and bandwidth. The key-value (KV) cache grows with
context length and the number of concurrent requests, while dense attention
reads the KV cache at every decoding step. Its footprint therefore
limits how many requests can run together~\cite{pagedattention2023}, and its repeated reads can
limit decoding speed~\cite{oaken2025}. KV cache compression offers an
opportunity to ease both constraints by reducing the amount of data stored and
accessed during generation. To realize these benefits, attention must
read the compressed KV cache efficiently, and the serving system must be
able to use the saved memory for additional requests.

Compression methods also differ in whether they preserve the values in
the KV cache. Lossy compression methods, such as
quantization~\cite{kivi2024,kvquant2024,comet2025}, reduce storage by approximating
the values in the original KV cache. Other methods retain only selected
tokens in the KV cache~\cite{h2o2023,streamingllm2024}. Their effects on model quality
depend on the model and workload. In long-context retrieval and
question-answering experiments, Li et al.~\cite{qllmeval2024} find that most
tested models are more sensitive to KV cache quantization than to weight
quantization at the same bit width. We therefore focus on lossless KV
cache compression for deployments that require the values in the original
KV cache to be preserved.

\begin{figure}[t]
    \centering
    \includegraphics[width=\columnwidth]{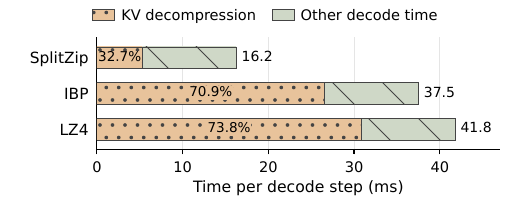}
    \caption{Instrumented decode time for lossless KV cache codecs adapted
    to SGLang for Llama 3.1 8B, with BF16, batch size 32,
    2048 input tokens. Each bar averages six repetitions
    of 256 decoding steps. Percentages show the decompression share.
    Prefill and initial compression are outside timing.}
    \label{fig:decompression-breakdown}
\end{figure}

KV cache transfers arise both in offloaded inference~\cite{flexgen2023}
and in serving that separates prefill from decoding~\cite{distserve2024,splitwise2024}.
Prior work uses lossless KV cache compression to reduce these transfer costs.
IBP targets transfers between CPUs and GPUs, while SplitZip targets
transfers between servers that process prompts and those that generate
outputs~\cite{ibp2026,splitzip2026}. These approaches reduce the bytes
transferred, but attention still reads the uncompressed KV cache from GPU
memory after reconstruction. The repeated KV cache reads that can make
decoding memory-bound therefore remain unchanged. Decompression also requires
execution time and GPU memory for the restored values.
Figure~\ref{fig:decompression-breakdown} measures this cost for three
lossless codecs adapted to SGLang, each decompressing the KV cache before
attention. Decompression accounts for 32.7\% to 73.8\% of instrumented
decode time in this configuration. Using compressed
KV cache directly for attention can reduce the bytes read at each decoding
step. Reconstructing values within the attention kernel avoids a separate
decompression pass and the need to store its uncompressed output in GPU
memory. The space released by compression can then be used to admit
additional requests.

We present \textsc{\sysname{}}, which enables attention to consume
the losslessly compressed KV cache directly without maintaining an
uncompressed copy in GPU memory. Completed KV cache pages remain
compressed in GPU memory. Attention reads the compressed KV cache blocks
and reconstructs the values it needs directly in registers. The KV cache
manager reports the additional
GPU memory made available by compression to the scheduler, allowing it
to admit more requests.
Keeping the KV cache compressed reduces data movement both between
devices and during attention, lowers the KV cache memory footprint, and eases the
memory bandwidth bottleneck during decoding.

Realizing these benefits requires addressing two challenges. First,
reducing decode latency requires a lossless compression format that
attention can use efficiently. The time saved by reading fewer bytes
can otherwise be offset by the work needed to reconstruct values from
the KV cache.
We design the compression format and attention kernel together,
arranging compressed keys and values to match their respective matrix
computations and reconstructing values directly in registers with simple
bit operations. Second, the memory saved by compression is needed both
to serve additional requests and to accommodate KV cache growth for
requests already running. The KV cache manager tracks how much space is
available for KV cache writes as generation and compression proceed.
The scheduler uses this information to admit additional requests while
reserving space for
ongoing generation.

We implement \sysname{} for the KV cache in the 16-bit bfloat16 (BF16) format,
on the top of FlashInfer~\cite{flashinfer2025} for attention and SGLang~\cite{sglang2024} for the KV cache manager.
We evaluate Llama 3.1 8B Instruct and Qwen3-8B on an NVIDIA H100.
Across the evaluated configurations, \sysname{} achieves $1.23\times$
to $1.25\times$ speedup over FlashInfer in decode attention.
With 8K inputs, it increases
model decode throughput by 26.3\% for Llama and 25.8\% for Qwen
over uncompressed SGLang. Each system uses its largest supported batch
under the same GPU memory budget.
Relative to the best-performing lossless KV cache compression
baseline, it delivers $2.76\times$ throughput for Llama and
$2.61\times$ for Qwen.

\section{Background and Motivation}
\label{sec:background}

\subsection{KV Cache and BF16}
\label{sec:background-kv}

LLM inference begins with prefill, which processes the input prompt and
builds the KV cache. Decoding then generates one token per request at
each step and extends the KV cache. Dense attention compares the current
query with stored keys to compute attention weights, then
combines stored values using these weights. The same portion of the KV
cache can therefore be read many times after it is written. Longer
contexts increase these reads and the memory occupied by the KV
cache~\cite{oaken2025}. Serving systems commonly organize the KV cache
into pages, each covering a fixed
number of tokens, and allocate more pages as requests grow~\cite{pagedattention2023}.

We consider the KV cache stored in BF16, a 16-bit floating-point format.
Each value contains one sign bit, eight exponent bits, and seven fraction
bits. Lossless compression preserves the original bit pattern of every
value, including all three fields.

\subsection{Compression for Faster Decode Attention}
\label{sec:background-compression}

Efficient attention kernels reduce GPU memory traffic through
tiling~\cite{flashattention2022} and improve execution through better
work partitioning~\cite{flashattention22024} and reduced softmax
synchronization~\cite{flashdecoding2024}.
KV cache compression can further reduce the bytes that these kernels read.

Lossless compression exploits repeated exponents or other bit patterns
in the KV cache. IBP and SplitZip use this redundancy to reduce transfer
costs~\cite{ibp2026,splitzip2026}. The repeated use of the KV cache creates
a further opportunity. Once a completed KV cache block is compressed,
subsequent decoding steps can read fewer bytes each time they consume it.
Restoring the KV cache after transfer leaves these repeated reads
uncompressed. Keeping it compressed for attention instead allows one
encoding operation to benefit many decoding steps. This makes attention
execution an important target for lossless KV cache compression.

Keeping the KV cache compressed also makes reconstruction part of the
decoding cost. Figure~\ref{fig:decompression-breakdown} examines this
cost in a common SGLang decode path. We run Llama 3.1 8B with 2048
input tokens and batch size 32 on one H100. SplitZip, IBP, and
nvCOMP LZ4~\cite{nvcomp}
each retain the compressed prompt KV cache and restore one layer into
a shared buffer before FlashInfer attention. We measure six repetitions
of 256 decoding steps, with prefill and initial compression outside timing.

Decompression takes 5.3 to 30.8\,ms per step, while the remaining
instrumented time stays near 10.9\,ms across the three codecs. The
decompression intervals sum K and V reconstruction across all layers.
The remaining time includes
attention and other model work. Attention still reads the restored
KV cache from GPU memory, so separate reconstruction adds work without
reducing these repeated reads. The measurement identifies an execution
cost that compression ratio alone does not capture. For lossy KV cache
compression, Gao et al.~\cite{rethinkingkv2025} find that throughput
benefits depend on batch size and prompt length.

Shannon entropy provides a reference for how compactly symbols can be
encoded on average~\cite{shannon1948},
but does not determine how quickly attention can locate and reconstruct
the values it needs.
Reconstruction within attention avoids that intermediate buffer, but its
instruction cost still affects latency. The compression format therefore
needs to support inexpensive reconstruction and arrange keys and values
for their respective attention computations. Designing the format and
attention together offers a way to retain the bandwidth savings while
limiting the work needed to use the compressed KV cache.

\subsection{Compression for Higher Serving Capacity}
\label{sec:background-capacity}

Compression can also make room for additional requests within the same
GPU memory budget. The available space changes as completed KV cache
pages are compressed and running requests extend their KV cache.
Scheduling based on a fixed capacity derived from uncompressed page
sizes cannot use this additional space. Yet admitting more requests
requires accounting for both their KV cache and the space needed by
ongoing generation. The relevant capacity is therefore the physical
space available for further KV cache writes. Exposing this capacity to
the scheduler connects compression to request admission. More concurrent
requests also increase the amount of KV cache read at each decoding
step~\cite{oaken2025}, linking this capacity opportunity to the need for
efficient attention over the compressed KV cache.

\section{\sysname{} Design}
\label{sec:methodology}

\subsection{Overview}
\label{sec:design-overview}

\sysname{} enables attention to consume compressed KV cache directly.
A KV cache page groups data for a fixed number of tokens. Within each
layer, the page's keys and values are divided into smaller compression
blocks that are encoded independently. GPU high-bandwidth memory (HBM)
is allocated in larger chunks, each of which can hold multiple pages.

\begin{figure}[t]
    \centering
    \includegraphics[width=\columnwidth]{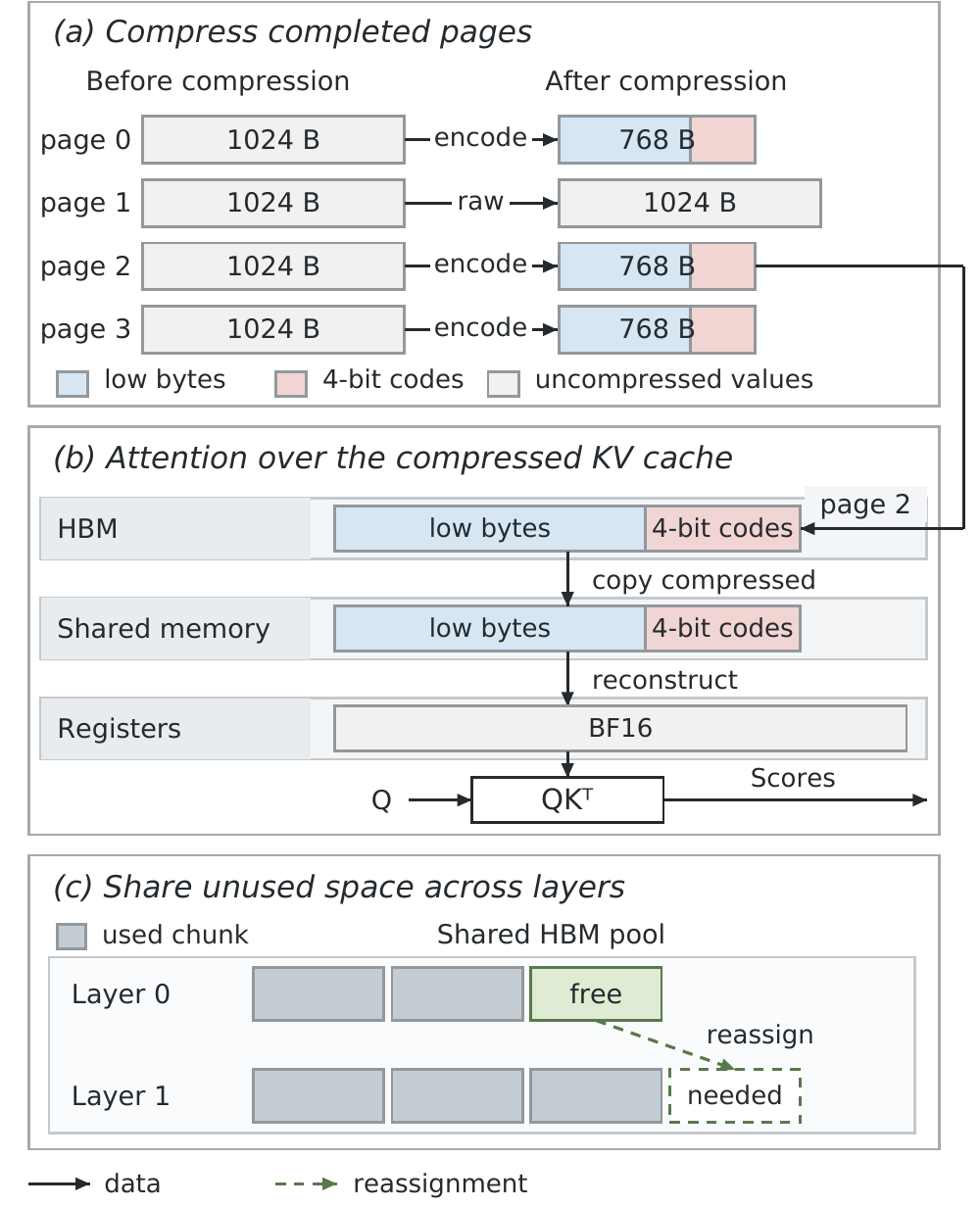}
    \caption{KV cache compression, attention, and memory allocation.
    (a) blocks that satisfy the compression condition shrink to 75\% of
    their original size, while the remaining blocks stay unchanged.
    (b) Data from the shown block of page 2 remain compressed through
    shared memory and are restored to BF16 in registers.
    (c) Each rectangle represents a memory chunk that can hold multiple
    pages. The empty chunk in layer 0 is reassigned to layer 1.}
    \label{fig:exactkv-overview}
\end{figure}

Figure~\ref{fig:exactkv-overview} shows how \sysname{} combines
KV cache compression, attention, and memory management.
Figure~\ref{fig:exactkv-overview}(a) illustrates page compression.
Within each page, blocks that satisfy the compression condition
shrink to 75\% of their original size, while the remaining blocks
stay unchanged. Attention reads these blocks in their stored
formats and restores compressed values in registers immediately
before computation, as shown in
Figure~\ref{fig:exactkv-overview}(b). This reduces memory reads
without writing the restored values back to GPU memory.
Compression also frees space for additional KV cache pages.
As shown in Figure~\ref{fig:exactkv-overview}(c), memory chunks
released by one layer can be reassigned to another layer,
allowing layers to share the memory saved by compression.

\subsection{Lossless KV Cache Format}
\label{sec:design-format}

\sysname{} divides the KV cache within each page into
blocks and compresses each block independently. When the exponents
within a block span a small range, we encode them as offsets from a
shared base while preserving every sign and fraction bit. Blocks that
do not meet this condition remain uncompressed.

\begin{figure*}[t]
    \centering
    \includegraphics[width=\textwidth]{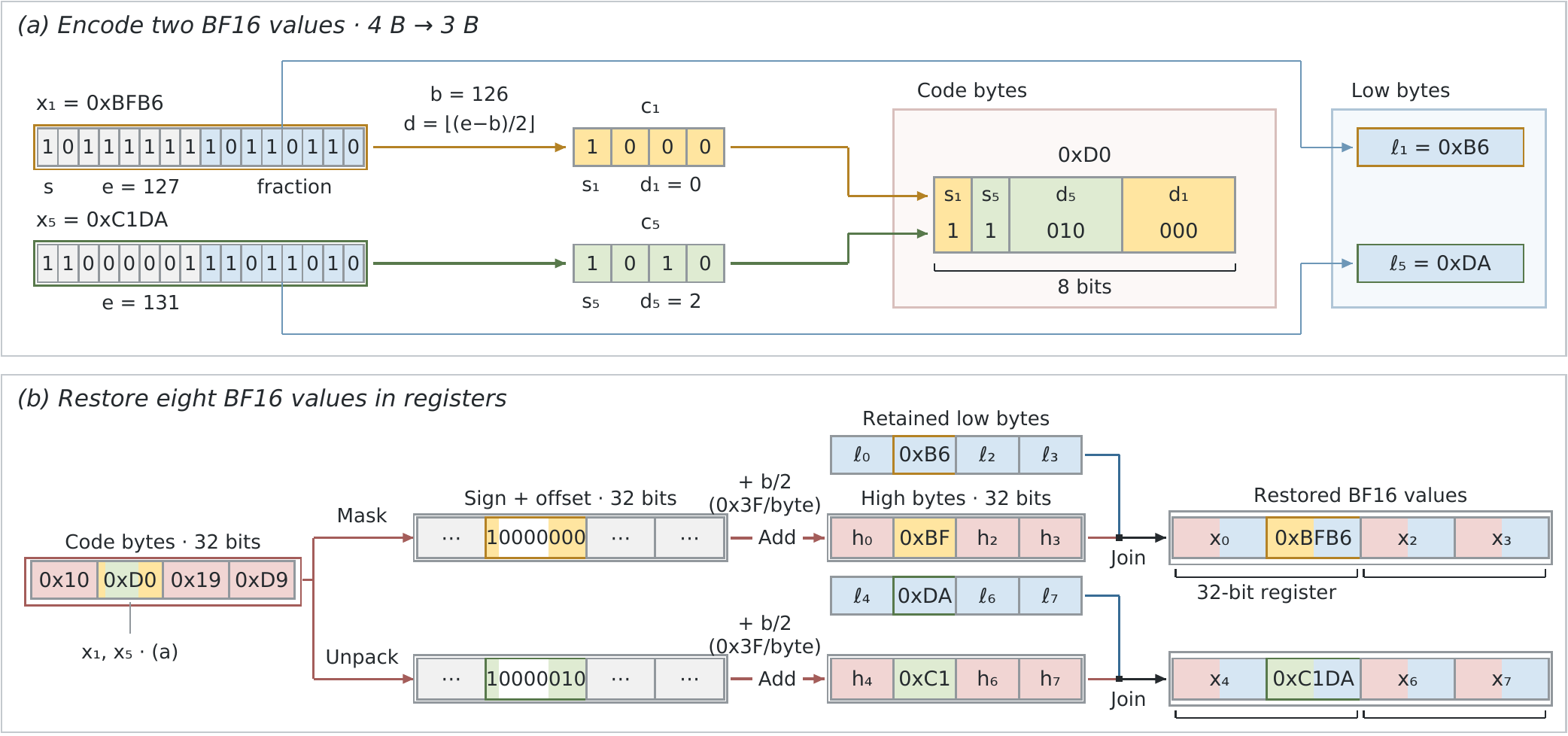}
    \caption{Lossless encoding and register reconstruction.
    Panel (a) follows two values from the eight-value group reconstructed
    by one thread in (b). Each branch in (b) restores four high bytes with
    one 32-bit addition. Join pairs each restored high byte with its
    retained low byte to recover the original BF16 value.
    Yellow and green track $x_1$ and $x_5$ across both
    panels. Subscripts identify the element to which each field belongs.
    Blue denotes retained low bytes $\ell_i$, and $h_i$ denotes the high
    byte of $x_i$. The prefix \texttt{0x} marks hexadecimal values;
    bit strings are binary. Data byte counts exclude metadata.}
    \label{fig:exactkv-format}
\end{figure*}

Let $s$, $e$, and $f$ denote the unsigned sign, exponent, and fraction
fields of a BF16 value. Here, $e$ is the unsigned integer stored in the
eight-bit exponent field. The base $b$ is the minimum exponent in
the block rounded down to an even integer. A block is compressed when
every exponent $e$ lies in $[b,b+15]$. Otherwise, it remains uncompressed.

We retain the low byte $\ell$ of each BF16 value unchanged.
This byte contains all seven fraction bits and the lowest exponent
bit. For a compressed block, the exponent offset $e-b$ fits in
four bits. Because $b$ is even, the lowest bit of this offset is
already preserved in $\ell$. We therefore store only its upper
three bits,
\begin{equation}
    d = \left\lfloor \frac{e-b}{2} \right\rfloor.
    \label{eq:kv-encode}
\end{equation}
For each value, we combine the sign bit $s$ and the three bits of
$d$ into a four-bit code, denoted by $c$. Its highest bit stores
$s$, and its lower three bits store $d$.
Each value thus requires one retained byte
and one four-bit code, or 12 bits in total. Two codes are packed
into one byte, as illustrated in Figure~\ref{fig:exactkv-format}(a).
An eligible block therefore occupies 75\% of its original size,
excluding metadata.

To restore the exponent, we read the three-bit offset $d$ from
the code $c$ and the retained exponent bit $r$ from the highest
bit of the low byte $\ell$. The original exponent is then
\begin{equation}
    \hat e = b + 2d + r.
    \label{eq:kv-reconstruct}
\end{equation}
The sign is recovered from the highest bit of $c$, and the
seven fraction bits are retained in $\ell$. All three fields
are therefore restored exactly.
These integer operations on bit fields introduce no floating-point
rounding.

For example, consider a 16-token page with a head dimension of 128.
For each KV head, the page holds $16\times128=2048$ elements for keys
and the same number for values. Dividing each set into 512-value
blocks gives four blocks for keys and four for values. Each eligible
block then occupies 768 bytes instead of 1024 bytes, excluding metadata.

Within each compressed block, we store the retained low bytes
contiguously, followed by the packed four-bit codes.

To locate blocks within a page, attention also needs their
starting positions and sizes. For each KV head, we store
separate metadata for the four key blocks and the four value
blocks. Each group uses eight bytes of metadata. Four bytes
store the starting offset of the first block, and the remaining
four bytes describe the blocks individually. Each description
either marks a block as uncompressed or stores its exponent
base divided by two, $b/2$. This identifies whether the block
occupies 1024 or 768 bytes and supplies the base needed for
reconstruction. Attention locates each subsequent block by
adding the sizes of the preceding blocks to the starting offset.

\subsection{Attention over the Compressed KV Cache}
\label{sec:design-attention}

Attention computes scores $QK^{\mathsf{T}}$ from queries $Q$ and keys $K$, then
applies scaling and softmax to obtain attention probabilities $P$.
The second matrix operation, $PV$, computes a weighted sum
of the values $V$ using the attention probabilities $P$.

Tensor Core matrix instructions require each thread to hold
specific matrix elements in its registers. We arrange the
KV cache to match this requirement, storing eight elements
needed by the same thread together within each compression block.
Keys and values use different layouts because they serve as
different operands in $QK^{\mathsf{T}}$ and $PV$.
The KV cache write kernel stores elements in these layouts,
and compression preserves the grouping. For each compressed group,
a thread reads eight retained low bytes and four code bytes,
then reconstructs the eight BF16 elements in its registers
for the Tensor Core instruction.

Only the high bytes, which contain the sign and upper seven
exponent bits, need to be reconstructed because the low bytes
are retained unchanged. We pack the codes to simplify this
reconstruction in registers. Each code byte combines the codes of two values
from an eight-element group.
Figure~\ref{fig:exactkv-format}(a) illustrates the pair
$x_1$ and $x_5$. We place the sign of $x_1$ in bit 7 and its
three-bit offset in bits 2 through 0, leaving the middle
four bits for the code of $x_5$, producing the byte \texttt{0xD0}
in the example. This places the sign and
offset of $x_1$ in the positions needed to reconstruct its
high byte. The sign and offset of $x_5$ can be moved into
the same positions with bit operations. The other three
code bytes follow the same pattern for $(x_0,x_4)$,
$(x_2,x_6)$, and $(x_3,x_7)$.

With this packing, the thread restores four high bytes at a time,
as shown in Figure~\ref{fig:exactkv-format}(b).
The four code bytes form one 32-bit word. In the upper
branch, masking keeps the signs and offsets of $x_0$ through $x_3$.
In the lower branch, unpacking moves those of $x_4$ through $x_7$ into
the same bit positions. Each branch then adds a 32-bit word with $b/2$
in each byte. This restores four high bytes with one integer addition.
The highlighted paths restore \texttt{0xBF} for $x_1$ and
\texttt{0xC1} for $x_5$ from the packed byte \texttt{0xD0}.
The exponent sum fits below the sign bit, so no carry changes the
sign or crosses into another byte. The restored high bytes combine
with the eight retained low bytes to form eight BF16 values, packed
into four 32-bit registers for the matrix operation.

The attention kernel reads the metadata and copies KV cache blocks
from HBM into on-chip shared memory. Compressed blocks remain compressed
during this copy. Threads read their groups from shared memory and restore
the values in registers immediately before use.
Figure~\ref{fig:exactkv-overview}(b) shows this path for keys. Values
follow the same steps before $PV$, and uncompressed blocks supply
the same operands without reconstruction. The kernel fetches metadata
for the next portion of the KV cache while computing attention on the
current portion. Attention consumes the compressed blocks without a
separate decompression pass or an uncompressed output buffer in GPU
memory.

\subsection{KV Cache Management and Scheduling}
\label{sec:design-scheduling}

\sysname{} keeps partially filled KV cache pages uncompressed
so that each decoding step can append new values directly.
Once a page is full, it can be compressed after the current
attention reads finish.
Figure~\ref{fig:exactkv-overview}(a) illustrates the compressed and
uncompressed blocks that result. The encoded block sizes determine the
allocation needed for each page. The page identifier stays the same
even if compression moves its data. A lookup table maps this identifier
to the page's current location in GPU memory. If a separate destination
is unavailable, compression can finish in place. Once compression
finishes, the page has no extra uncompressed copy.

Each new logical KV cache page needs physical space in every layer.
The number of pages that can be added therefore depends on the space
available across all layers. Each layer has private storage, and a
shared pool of memory chunks can provide additional space. Once a
chunk from this pool is empty and outstanding reads have finished, it can
return to the pool and be reassigned to another layer, as shown in
Figure~\ref{fig:exactkv-overview}(c). This allows shared space released
in one layer to serve allocations in another.

The allocator determines how many additional KV cache pages
the remaining memory can support across all layers. It uses
the uncompressed page size because new pages must accept
writes before they can be compressed. As compression releases
memory, the allocator updates the available capacity.
The updated capacity allows more requests to be served while
preserving space for ongoing generation.

\section{Evaluation}
\label{sec:evaluation}

\subsection{Experimental Setup}
\label{sec:eval-setup}

\noindent\textbf{Models.}
We evaluate \sysname{} with Llama 3.1 8B Instruct~\cite{llama31model}
and Qwen3-8B~\cite{qwen3model}. Weights and the original KV cache
use BF16 and reside on a single GPU.

\noindent\textbf{Testbed.}
We use an NVIDIA H100 with 80\,GB of HBM3, CUDA 13.0,
PyTorch 2.11.0, and FlashInfer 0.6.12. \sysname{} is integrated
into SGLang with 16-token KV cache pages. All framework comparisons
enable CUDA Graphs.

\noindent\textbf{Baselines.}
We compare against uncompressed SGLang~\cite{sglang2024} with
FlashInfer~\cite{flashinfer2025} attention and three lossless codecs,
IBP~\cite{ibp2026}, SplitZip~\cite{splitzip2026}, and nvCOMP
LZ4~\cite{nvcomp}. Their public implementations are integrated into
the same SGLang decode path. Each codec decompresses a layer's
KV cache into a GPU buffer before FlashInfer attention, while
\sysname{} reconstructs values within attention.

\noindent\textbf{Workloads and measurements.}
Framework experiments use fixed essay prompts with 8192 input
tokens and 256 output tokens per request. We time the 255 decode
steps after the first token, excluding prefill and initial compression.
Newly generated KV cache remains uncompressed for all methods.
Throughput is total tokens generated during these steps divided by
total decode time across runs. We verify KV cache reconstruction bitwise.

\subsection{Attention Kernel Performance}
\label{sec:eval-kernel}

\begin{figure}[!tb]
    \centering
    \includegraphics[width=\columnwidth]{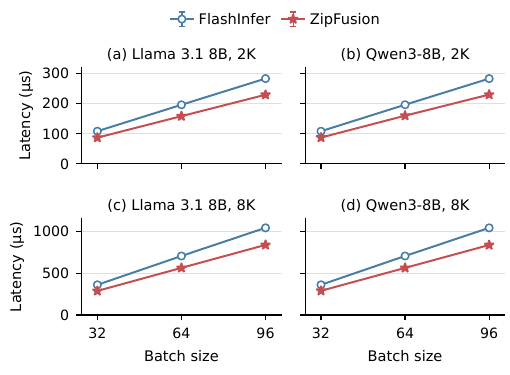}
    \caption{Decode attention latency.
    Points show medians of 400 samples.}
    \label{fig:attention-performance}
\end{figure}

We benchmark attention using queries and KV cache captured from
a middle layer of each model, with separate KV cache copies for
each batch entry. Figure~\ref{fig:attention-performance} compares
2048- and 8192-token contexts at batch sizes 32, 64, and 96.
We compare \sysname{} against production FlashInfer.
L2 is cleared before each timed call. Timing includes attention,
fused reconstruction, and any split-result merge, excluding
input preparation and initial compression.

\begin{figure*}[!t]
    \centering
    \includegraphics[width=\textwidth]{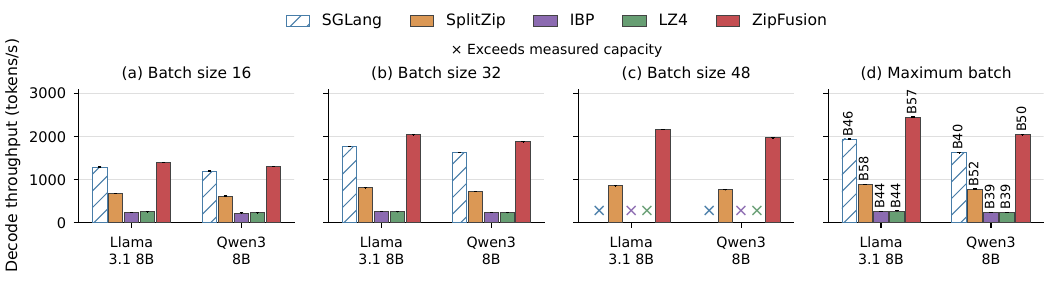}
    \caption{Model decode throughput with precompressed 8192-token
    prefixes. Panel (d) labels each method's largest supported batch.
    Whiskers show the minimum and maximum over six runs.
    Crosses mark batches above measured capacity.}
    \label{fig:framework-throughput}
\end{figure*}

Across the evaluated configurations, \sysname{} achieves
$1.23$--$1.25\times$ speedup over production FlashInfer,
corresponding to 18.5\%--20.2\% lower attention latency.
The improvement holds for both models at both context lengths and
all three batch sizes. At an 8192-token context and batch size 64,
Llama attention falls from 704.6 to 562.7\,$\mu$s, while Qwen
falls from 704.8 to 562.5\,$\mu$s. These timings include
reconstruction, showing that its cost is small enough to retain
a net latency benefit from compressed attention.

\subsection{Framework Decode Throughput}
\label{sec:eval-framework}

Figure~\ref{fig:framework-throughput} compares fixed batch sizes
of 16, 32, and 48, and each method's largest supported batch.
All methods use a memory fraction of 0.8, including compression
metadata and temporary buffers in the memory budget. At batch size 32,
\sysname{} reaches 2044 tokens/s for Llama and 1882 tokens/s for
Qwen, improving on uncompressed SGLang by 15.3\% and 15.2\%,
respectively, and achieves the highest throughput among the
evaluated methods at this batch size.

Relative to SplitZip, the fastest compressed baseline at batch size 32,
\sysname{} delivers $2.51\times$ throughput for Llama and
$2.57\times$ for Qwen. These gains hold with the same number of
requests in each batch. By reconstructing values within attention,
\sysname{} avoids the separate decompression pass and reads less
KV cache data from GPU memory during attention.

The saved memory also supports larger batches within the same budget,
as shown in Figure~\ref{fig:framework-throughput}(d).
\sysname{} increases the supported batch size from 46 to 57 for
Llama and from 40 to 50 for Qwen. Compared with SGLang at its own
largest batch, throughput improves by 26.3\% and 25.8\%, respectively.
\sysname{} reaches 2453 tokens/s for Llama and 2042 tokens/s for
Qwen, about $1.26\times$ the throughput of SGLang in both cases.
The larger gains at maximum batch size combine faster attention
with the capacity to serve more requests within the same memory budget.

\section{Related Work}
\label{sec:related-work}

IBP~\cite{ibp2026} and SplitZip~\cite{splitzip2026} use lossless KV cache
compression to reduce transfers between CPUs and GPUs or between prefill
and decode workers. Both restore the KV cache before attention.
\sysname{} keeps completed pages compressed during attention, reducing
their footprint and repeated GPU memory reads.

For weights, NeuZip~\cite{neuzip2024} and DFloat11~\cite{dfloat112025}
exploit the low entropy of exponents. Tan et al.~\cite{shannon2026} align entropy decoding with
matrix multiplication tiles. ZipServ~\cite{zipserv2026} fuses weight
decompression with matrix multiplication, restoring operands in registers.
ZipServ emphasizes GDDR GPUs and reports that reconstruction overhead
can offset bandwidth savings on A100 and H800. \sysname{} targets the
growing KV cache and demonstrates benefits on an H100 with HBM3.
For 8192-token inputs with precompressed KV cache prefixes, it achieves
up to 15.3\% higher model decode throughput than uncompressed
SGLang at batch size 32, and up to 26.3\% when each system uses its
largest supported batch under the same GPU memory budget
(Section~\ref{sec:eval-framework}).

Orca~\cite{orca2022} schedules requests at each generation step, while
Sarathi-Serve~\cite{sarathi2024} chunks prefill work to limit interference
with decoding. vAttention~\cite{vattention2025} allocates physical memory
on demand while keeping the KV cache contiguous in virtual memory.
\sysname{} connects memory reclaimed by compression to request admission
while reserving space for ongoing generation.

\section{Conclusion}
\label{sec:conclusion}

We presented \sysname{}, which co-designs lossless KV cache compression,
attention, and page management to reduce KV cache memory use and accelerate
decoding. Attention reconstructs KV cache values directly in registers,
avoiding a separate decompression pass and an uncompressed output buffer
in GPU memory. The page manager makes saved memory available to the
scheduler while reserving space for continued generation. On an NVIDIA
H100, \sysname{} reduces attention latency by 18.5\% to 20.2\% relative
to FlashInfer in the evaluated configurations. With 8K input contexts
and precompressed KV cache prefixes, \sysname{} improves model
decode throughput by up to 26.3\% over uncompressed SGLang. Each system
uses its largest batch within the same memory budget. These results
show that lossless KV cache compression can support both larger
batches and faster decoding.

\bibliographystyle{IEEEtran}
\bibliography{refs}

\end{document}